\documentclass[pra,onecolumn,showpacs,amsmath,amssymb,pra,floatfix]{revtex4}
\usepackage{graphicx}% Include figure files
\usepackage{dcolumn}% Align table columns on decimal point
\usepackage{bm}% bold math
\begin{document}

%\preprint{APS/123-QED}

\title{Electromagnetic transitions of the helium atom in superstrong magnetic fields}
\date{\today}

\author{Omar-Alexander Al-Hujaj}
\email{Alexander.Al-Hujaj@pci.uni-heidelberg.de}
\affiliation{%
Theoretische Chemie, Institut f\"ur Physikalische Chemie der Universit\"at Heidelberg, INF 229, 69120 Heidelberg, Germany}%
\author{Peter Schmelcher}
\email{Peter.Schmelcher@pci.uni-heidelberg.de}
\affiliation{%
Theoretische Chemie, Institut f\"ur Physikalische Chemie der Universit\"at Heidelberg, INF 229, 69120 Heidelberg, Germany}%
\affiliation{%
Physikalisches Institut der Universit\"at Heidelberg, Philosophenweg 12, 69120 Heidelberg, Germany}%

\date{\today}% It is always \today, today,

\begin{abstract}
  We investigate the electromagnetic transition probabilities for the
  helium atom embedded in a superstrong magnetic field taking into
  account the finite nuclear mass.  We address 
  the regime $\gamma=100-10\ 000$~a.u. studying several excited states
  for each symmetry, i.e. for the magnetic 
  quantum numbers $0,-1,-2,-3$, positive and negative z~parity and
  singlet and triplet symmetry. The oscillator strengths as a function
  of the magnetic field, and in particular the influence of the finite
  nuclear mass on the oscillator 
  strengths are shown and analyzed.
\end{abstract}

\maketitle
\section{Introduction}
Exposing matter to strong and superstrong magnetic fields (which are
fields of the order of $10^5$~T and above) dramatically
changes its properties and yields new and unexpected phenomena. On the
microscopic scale, i.e. for atomic and molecular systems, magnetic
forces have a tremendous influence on the electronic structure and
quantum dynamics
\cite{Friedrich:1989_1,Ruder:1994_1,Schmelcher:1998_1}. This is due to
the 
different appearances 
of the Coulomb and 
magnetic forces. From a theoretical point of view, strong and
superstrong magnetic fields are interesting, because the competing
 forces prevent a perturbative
treatment of the problem. Therefore it is necessary to develop and
apply new nonperturbative techniques.

Certain astrophysical objects possess strong and
superstrong magnetic
fields~\cite{Angel:1978_1,Pavlov:1995_1,Wickramasinghe:2000_1}.
Atmospheres of magnetic white dwarfs are 
exposed to  fields of the order of 100~--~10$^5$~T, magnetic fields in
the photosphere of
neutron stars  are of the order of
10$^5$~--10$^{10}$~T. For the interpretation of the spectra of 
these astrophysical objects a wealth of highly accurate atomic and molecular
energies, transition wavelengths and transition probabilities are
needed. An example for the analysis of astrophysical spectra of
magnetized objects using atomic data in strong fields is the white dwarf
GrW$+70^\circ 8247$, which represents a cornerstone for the
understanding of magnetic 
white dwarfs in general \cite{Angel:1985_1,Greenstein:1985_1,Wunner:1985_1,Wickramasinghe:1988_1}. 

Highly accurate data are available for hydrogen in strong magnetic
fields since more than a decade 
\cite{Friedrich:1989_1,Ruder:1994_1,Kravchenko:1996_1}. This system is
now understood to a very 
high degree. However beyond hydrogen, there is significant interest in
detailed data on heavier elements, such as He, Na, Fe and even
molecules. Especially helium plays an important role in the
atmospheres of magnetic white dwarfs and potentially also neutron stars.
The electronic structure of the helium atom has been considered by several authors during the last decades
\cite{Mueller:1975_1,Virtamo:1976_1,Proeschl:1982_1,Vincke:1989_1,Braun:1993_1,Ivanov:1994_1,Thurner:1993_1,Braun:1998_1,Heyl:1998:_1}.
However most of the corresponding investigations are restricted to a few states or field strengths. Only a
few works provide accuracies, that are necessary for astrophysical applications.

Recently  detailed investigations of helium in the strong field
regime have been performed,  providing the community with detailed
energy levels, transition wavelengths, and transition probabilities, for
a dense grid of field strengths in the range of $0\leq\gamma\leq100$~a.u. (one atomic units
corresponds to $2.3505$~$\times$~$10^5$~T)
\cite{Becken:1999_1,Becken:2000_1,Becken:2001_1,Becken:2002_1}.  Numerous
 symmetries and many excited states have been addressed. With the
 resulting large  amount of
data it was possible to identify the  absorption edges of the
observational spectrum of the magnetic white dwarf GD229 \cite{Green:1980_1,Schmidt:1990_1,Schmidt:1996_1}, which have been unexplained for more than
25 years \cite{Jordan:1998_1,Jordan:2001_1}. 

At this point also the work by Jones et. al \cite{Jones:1999_1} should
be mentioned. They applied a released-phase quantum
Monte Carlo method 
in order to evaluate bound state energies and dipole-matrix elements
for the ground and a few excited 
triplet states. This has been done for a grid of several field
strengths $0.08\leq\gamma\leq800$~a.u. 

Addressing the superstrong field regime a further challenge is the problem of the finite nuclear mass. The
dominant energy correction, caused by the finite nuclear  mass is for
a  field of $10^9$~T of
 the same order of magnitude as
the binding energy itself. This holds even for the energetically
lowest  states \cite{Al-Hujaj:2003_1}. Therefore 
effects due to the finite nuclear mass have to be taken 
into account for a correct description of the structure of the atom. 
Up to date there are no detailed studies about the influence of the
finite nuclear mass on the transition rates.

The purpose of the present article is to provide results on the transition
probabilities 
for helium in the superstrong field regime. In
Sect.~\ref{sec:electroMagTrans} we  review the expressions for
the transition matrix elements and analyze  the influence
of the finite nuclear mass. In Sect.~\ref{sec:results} we 
provide our results and discuss some particular features of the
transition probabilities as a function of the  field
strength.  Sect.~\ref{sec:concl} provides a brief conclusion and an outlook.

\section{Electromagnetic transition probabilities for finite nuclear
  mass}
\label{sec:electroMagTrans}

A detailed comparison of theoretical and
observational spectra requires not only  the energies and transition
wavelengths, but also the corresponding oscillator
strengths.  Selection 
rules of allowed and forbidden transitions are of particular
importance. Our investigation focuses on the dominant electric
dipole transitions. We will shortly review the derivation of the corresponding
operators since there are modifications due to the presence of the magnetic
field as well as the finite nuclear mass.

Our starting point is the pseudo-separated Hamiltonian
\cite{Lamb:1952_1,Avron:1978_1,Johnson:1983_1,Schmelcher:1994_1} using
relative coordinates $\{\bm{r}_{i}\}$ for the electrons with respect to the
nucleus in atomic units:  

\begin{eqnarray}
  \label{eq:Ham_rel}
  H&=&\sum_i\left\{\frac{1}{2}\left(\frac{ \bm{P}}{M_A}
    +\bm{p}_i+\frac{1}{2}\bm{B}\times\bm{r}_i\right)^2-\frac{2}{|\bm{r}_i|}\right\}\\
&+&\frac{1}{2M_0}\left(\frac{M_0}{M_A}\bm{P}-\sum_j \bm{p}_j +
    \frac{1}{2}\sum_j\bm{B}\times\bm{r}_j\right)^2\\
&+& \frac{1}{|\bm{r}_1-\bm{r}_2|}
\end{eqnarray}

Here is $M_A$ the total mass of the atom, $\bm{P}$ denotes the
pseudo-momentum, and $\bm{B}$ the magnetic field vector.

On the other hand, we have the operator $H_{rad}$ describing the
interaction of the 
system with the electromagnetic radiation field $\bm{A}_{r}$, neglecting
quadratic terms in $\bm{A}_{r}$. It is given
in relative coordinates by
\begin{eqnarray}
  \label{eq:H_int}
  H_{rad}&=&\sum_i\left(\frac{1}{M_A}\bm{P}+\bm{p}_i+\frac{1}{2}\bm{B}\times\bm{r}_i\right)\bm{A}_{r}(\bm{r}'_i)+\\
 && -\frac{2}{M_0}\sum_i\left(\frac{M_0}{M_A}\bm{P}-\sum_j\left(\bm{p}_j-\frac{\bm{B}\times\bm{r}_j}{2}\right)\right)\bm{A}_{r}(\bm{r}'_N).
\end{eqnarray}
Here $\bm{r}'_i$ denotes the  position vector of electron
$i$ and $\bm{r}'_N$ the position of the nucleus in the laboratory
frame. The radiative part of 
the electromagnetic field $\bm{A}_{r}(\bm{r})$ reads in quantized form
(we consider only the creation of  photons):
\begin{equation}
 \bm{A}_{r}(\bm{r})=\sum_{\bm{k},\lambda}N(\bm{k})a^+_{\bm{k},\lambda}\bm{\epsilon}_{\bm{k},\lambda}\exp(i
  \bm{k} \bm{r}+i \omega t)
\end{equation}
$a^+_{\bm{k},\lambda}$ denotes the creation operator for a photon with 
wave vector $\bm{k}$ and wavelength $\lambda$.
$\bm{\epsilon}_{\bm{k},\lambda}$ is the polarization vector of the
photon, whereas $N$ is an
amplitude. 
In the next step, we
will 
integrate over the center of mass coordinate $\bm{R}$ by calculating
the matrix element  of $H_{rad}$ between
two eigenfunctions of the pseudo-momentum $\bm{P}$ (eigenvalues are
denoted by $\bm{K_i}$ and $\bm{K_f}$), which are given by 
expressions of the form: 
\begin{equation}
  \frac{1}{\sqrt{V}}\exp\left(-i \bm{K}\cdot\bm{R}\right)
\end{equation}
if we assume an integration volume $V$.

The dipole approximation, which reads
$\exp(i\bm{k}\cdot\bm{r}_i)\approx1$, leads us in
first order time dependent perturbation theory  to the following
expression for the transition 
rates:
\begin{eqnarray}
  \frac{d
  P_{fi}}{dt}&=&2\pi\sum_\sigma\left[\delta(E_f-E_i-\omega)\delta_{\bm{K}_i,\bm{K}_f-\bm{k}}\times\right.\\
&&\left.\left|<i|\bm{G_\sigma^+}|f>\right|^2\right]
\end{eqnarray}
where
\begin{equation}
  \bm{G_\sigma^+}=-N(\bm{k})a^+_{\sigma}\bm{\epsilon}_\sigma^*\sum_i\left(\frac{M_A}{M_0}\bm{p}_i+\frac{M_0-2}{2M_0}\bm{B}\times\bm{r}_i\right),
\end{equation}
and $<i|$, $|f>$, denote the electronic initial and final states,
respectively. In the following we will assume that the wavevector
$\bm{k}$ is much smaller than $\bm{K}_i,\bm{K}_f$, which is
well-justified in atomic transitions.
 Thus using
$\bm{\epsilon}_\sigma^*(\frac{M_A}{M_0}\bm{p}+\frac{M_0-2}{2M_0}\bm{B}\times\bm{r})=:Q_\sigma$,
we obtain the following expressions for the electronic transitions:
\begin{eqnarray}
\label{eq:transrate_vel}
  p_{fi}^{\sigma}&=&\frac{2}{E_f-E_i}<f|Q_\sigma|i>,\\
  d_{fi}^{(\sigma)}&=&\left(\frac{2}{E_f-E_i}\right)^2|<f|Q_\sigma|i>|^2,\\
  f_{fi}^{(\sigma)}&=&\frac{E_f-E_i}{2}d_{fi}^{(\sigma)}.
\end{eqnarray}
These expressions represent the dipole-matrix element, the dipole
strength and the oscillator strength, respectively in the velocity
representation. 

On the other hand we have for the expectation value of
the commutator
\begin{eqnarray}
  <i|[H,\bm{r}]|f> &=&
  <i|\frac{M_A}{M_0}\bm{p}+\frac{M_0-2}{2M_0}\bm{B}\times\bm{r}|f>\label{eq:comm1}\\
&=&(E_i-E_f)<i|\bm{r}|f>,\label{eq:comm2}
\end{eqnarray}
where $\bm{r}:=\bm{r}_1+\bm{r}_2$ and $\bm{p}:=\bm{p}_1+\bm{p}_2$ are
symmetrized one-particle operators.
Applying the identity of Eqs.~(\ref{eq:comm1}),(\ref{eq:comm2}) we arrive at
the length representation, that reads, 
\begin{eqnarray}
  p_{fi}^{\sigma}&=&2<f|D_\sigma|i>,\\
  d_{fi}^{(\sigma)}&=&4|<f|D_\sigma|i>|^2,\\
  f_{fi}^{(\sigma)}&=&\frac{E_f-E_i}{2}d_{fi}^{(\sigma)},\label{eq:osc_str_len}
\end{eqnarray}
where $\bm{\epsilon}_\sigma^*\bm{r}=:D_\sigma$.
These above two representations are equivalent. However in case of numerical
calculations, the two representations  yield in general different results. The
relative deviation between the two representations is a good measure
for the
convergence of the computational method.  Only results that obey
certain consistency criteria concerning the length to velocity
representations of the transition rates are presented. This
ensures in particular the gauge independence of our results.

In the following we will assume a vanishing pseudo-momentum $\bm{K}$,
 which is an appropriate approximation in case of slow moving atoms.
 The basic polarization vectors $\bm{\epsilon}_\sigma$ are chosen to
 be  parallel and perpendicular to the magnetic field vector, 
indicated as components $z$ and $x\pm i y$. This
leads to the following selection rules for the electromagnetic
transitions of the helium atom in a magnetic field \cite{Becken:2000_2}:
\begin{eqnarray}
\label{eq:circ}
  |M_f-M_i|=1 & \mbox{and} & \Pi_{z_{f}}\Pi_{z_{i}}=1,
\end{eqnarray}
or 
\begin{eqnarray}
\label{eq:lin}
  |M_f-M_i|=0 & \mbox{and} & \Pi_{z_{f}}\Pi_{z_{i}}=-1
\end{eqnarray}
and
\begin{eqnarray}
  S_f-S_i=0 & \mbox{and} & S_{z_{f}}-S_{z_{i}}=0.
\end{eqnarray}
Here Eq.~(\ref{eq:circ}), (\ref{eq:lin})  describe circular and
linear polarized transitions, respectively.

To understand the influence of the finite nuclear mass we rewrite the
expression for the oscillator strength in the velocity form
Eq.~(\ref{eq:osc_str_len})   as 
\begin{equation}
  \label{eq:trans_len_sec}
  f_{fi}^{(\sigma)}=2\left(E_f-E_i\right)|<f|D_\sigma|i>|^2.
\end{equation}
The energy factor $(E_f-E_i)$ plays a significant role, as we will
see below. 
One result of Ref.\cite{Al-Hujaj:2003_1} is, that effects of
the mass polarization operators are small and therefore in a good approximation
results for finite nuclear mass can be expressed in terms of results
for infinite nuclear mass:
\begin{equation}
  \label{eq:redmass}
  E(M_0,\gamma)\approx \frac{1}{\mu'}E(\infty,\gamma)-\frac{\gamma
  M}{M_0}+\frac{2 \gamma}{M_0} \frac{\partial}{\partial \gamma}E(\infty,\gamma)
\end{equation}
Here $E(M_0,\gamma)$ denotes the total energy of an eigenstate for the
 Hamilton operator of the helium atom for nuclear
mass $M_0$ and a field strength $\gamma$. $\mu':=(1-1/M_0)^{-1}$ is a
reduced mass. First we will concentrate on
transitions which do not involve tightly bound states. For the
 corresponding transitions
 the last, i.e. third term on the right hand side of Eq.~(\ref{eq:redmass})
in general cancels in the energy 
factor of Eq.~(\ref{eq:trans_len_sec}).
As a consequence, there are two generic cases (no tightly bound states
 involved) 
for the influence of the energy factor $(E_f-E_i)$: In the case of linear polarized
transitions, the magnetic quantum numbers $M_i$ and $M_f$ are equal
and therefore 
 the energy factor is just scaled by the factor
$1/\mu'$, compared to the results for an infinite nuclear
 mass. Typically these oscillator strengths are approximately constant
 as a 
function of the field strength. Note that the factor $1/\mu'$ deviates
 for helium about $10^{-4}$ from 1.
In the case of
circular polarized transitions the second term on the r.h.s. of
Eq.~(\ref{eq:redmass}) $(-\gamma M/M_0)$ becomes important, since the
magnetic quantum numbers $M_i$ and $M_f$ are different.
Therefore the linear term $\gamma/M_0$ is added to the energy
factor, which in general causes an increase
 of the oscillator strengths compared to results for infinite
nuclear mass, of the form
 \begin{equation}
   f_{fi}^\sigma(M_0,\gamma)\approx
   \frac{1}{\mu'}f_{fi}^\sigma(\infty,\gamma)+ \frac{\gamma}{2 M_0}|p^\sigma_{fi}(\infty,\gamma)|^2.
 \end{equation}
However, the typical oscillator strengths for circular
 polarized transitions decrease according to a power law. 
We note that in case of transitions emanating from tightly bound states the
third term 
on the r.h.s. of Eq.~(\ref{eq:redmass}) becomes important and in general modifies
the pattern for linear polarized transitions. This is essentially due
 to the fact that the energies of
magnetically tightly bound states exhibit an inherently different
 field dependence than the corresponding quantity of 
non tightly bound states. The above discussed behavior will be
 observed when discussing our results of oscillator strengths in
 Sect.\ref{sec:results}.

Some comments on our computational approach are in order. The calculations
are performed using an anisotropic Gaussian basis set, which was put
forward by Schmelcher and Cederbaum \cite{Schmelcher:1988_1}, and 
which has been
successfully applied to several atoms, ions and
molecules\cite{Kappes:1996_1,Detmer:1997_1,Detmer:1998_1,Becken:1999_1,Becken:2000_1,Becken:2001_1,Becken:2002_1,Al-Hujaj:2000_1,Al-Hujaj:2003_1}.
The corresponding basis
functions have been optimized for each field strength to solve the  one particle
problems, i.e.  H and He$^+$ in a magnetic field. We refer the reader to Ref.\cite{Al-Hujaj:2003_1} for more details. It has been
shown, that this approach  yields  accurate energies and in particular
oscillator strength for helium,
by comparing with the corresponding data in the literature
\cite{Becken:1999_1,Becken:2000_1,Becken:2001_1,Becken:2002_1}. 

\section{Results}
\label{sec:results}

In this section, we present and discuss our results on the oscillator
strengths of electric dipole transitions of helium in the superstrong
field regime. In order to label the states, we use the standard
spectroscopic notation $n^{2S+1}M^{\Pi_z}$. Here $2S+1$ indicates the
spin multiplicity, $M$ is the magnetic quantum number, $\Pi_z$ the z~parity,
and $n$ the degree of excitation in the corresponding symmetry
subspace. The accuracy for the reported oscillator strengths is between 
$10^{-4}$ and a few times $10^{-2}$.

As discussed in Ref.\cite{Al-Hujaj:2003_1} the number of bound states
of the helium atom in superstrong
magnetic fields becomes finite, i.e. the spectrum terminates, if the
effects of the finite nuclear 
mass are taken into account. Therefore only a finite, usually small number of
transitions ``survive'' in the superstrong field regime. On the other
hand the ionization threshold (He~$\rightarrow$~He$^+$~$+$~e$^-$) is up
to date not known  exactly due to missing detailed investigations
on the moving He$^+$ ion in a magnetic field. The exact field strength
for which a certain state
becomes unbound is therefore unknown. Since our basis functions cannot
properly describe the electronic continuum we report here only on
transitions that are known to be energetically well-separated enough from
 the continuum.

\begin{figure*}
    \includegraphics[scale=0.5,angle=-90]{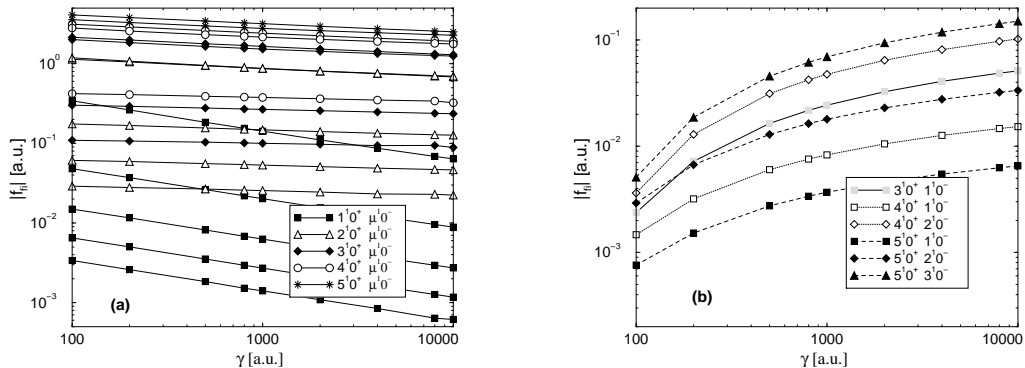}
\caption{\label{fig:osc_0s+sel}The absolute value of the oscillator
  strength $|f_{fi}|$  of the linear polarized transitions
    $\nu^10^+\rightarrow\mu^10^-$ as a function of the field
    strength $\gamma$. (a) On the left hand side from 
    bottom to top 
    $(\nu,\mu)$=$(5,4)$, $(5,5)$, $(4,3)$, $(4,4)$, $(3,2)$, $(3,3)$,
    $(2,2)$, $(2,1)$, $(4,5)$, $(1,1)$, $(3,5)$, $(2,4)$, $(1,2)$,
    $(2.5)$, $(1,3)$,$(1,4)$, $(1,5)$. (b) Oscillator strengths of a
    group of transitions belonging to 
    these symmetry subspaces showing a different field dependency.} 
\end{figure*}

The typical features of oscillator strengths of linear polarized
transition, discussed in Sect.\ref{sec:electroMagTrans}, can be clearly seen in
Fig.~\ref{fig:osc_0s+sel}~(a). The  oscillator strengths for
 several transitions stay constant, or change much less than one order
of magnitude. On the other hand transitions emanating from the tightly bound
state $1^10^+$ can be identified by their power law behavior. A
completely different pattern belongs to the transitions $3^1
0^+\rightarrow1^10^-$,$4^1 0^+\rightarrow1^10^-$,$4^1
0^+\rightarrow2^10^-$,$5^1 0^+\rightarrow1^10^-$,$5^1
0^+\rightarrow1^20^-$, and $5^1 0^+\rightarrow3^10^-$, depicted in Fig.~\ref{fig:osc_0s+sel}~(b). For a
field strength below a critical field strength $\gamma_c\approx50$ they
decrease (this can not be seen in Fig.~\ref{fig:osc_0s+sel}~(b)), above $\gamma_c$ they increase. Numerical
values for  transition wavelengths and oscillator strengths for a few
of the lowest 
linear polarized transitions  $\mu^{2S+1}
0^+\rightarrow\nu^{2S+1}0^-$, $\mu,\nu=1,2$ 
are presented in table  \ref{tab:numval_lin}.

\begin{table*}[htbp]
\begin{center}
\begin{ruledtabular}
\begin{tabular}{cddcdcdcdcccd}
%\hline
 &
 \multicolumn{2}{c}{$1^10^+\rightarrow1^10^-$}&\multicolumn{2}{c}{$2^10^+\rightarrow1^10^-$}&\multicolumn{2}{c}{$2^10^+\rightarrow2^10^-$}
 &\multicolumn{2}{c}{$1^30^+\rightarrow1^30^-$}&\multicolumn{2}{c}{$2^30^+\rightarrow1^30^-$}&\multicolumn{2}{c}{$2^30^+\rightarrow2^30^-$}\\
\cline{2-3}\cline{4-5}\cline{6-7}\cline{8-9}\cline{10-11}\cline{12-13}
\multicolumn{1}{c}{$\gamma$}     & \multicolumn{1}{c}{$\lambda$} & \multicolumn{1}{c}{$|f_{fi}|$} & \multicolumn{1}{c}{$\lambda$} & \multicolumn{1}{c}{$|f_{fi}|$} & \multicolumn{1}{c}{$\lambda$} & \multicolumn{1}{c}{$|f_{fi}|$} & \multicolumn{1}{c}{$\lambda$} & \multicolumn{1}{c}{$|f_{fi}|$}&\multicolumn{1}{c}{$\lambda$} & \multicolumn{1}{c}{$|f_{fi}|$}&\multicolumn{1}{c}{$\lambda$} & \multicolumn{1}{c}{$|f_{fi}|$}\\
\hline
100 &    88.546 &   0.3415  &   3916   & 1.133  & 3033  &1.186 & 1923  & 1.269 &1066  &$0.01508$&      12820&       2.50\\
200 &    69.545 &   0.2620  &    4089  & 1.05   & 2819  &1.08  & 2038  & 1.21  &1072  &$0.00685$ &     13710&       2.39\\
500 &    51.207 &   0.1843  &   4333   & 0.94   & 2586  &0.951 & 2255  & 1.12  &1092  &$6.3\times10^{-4}$ &     15340&       2.2\\
800 &    44.028 &   0.1544  &   4456   & 0.89   & 2485  &0.90  & 2389  & 1.070 &1104  &$2.1\times10^{-5}$&      16330&       2.10\\
1000 &   41.040 &   0.1421  &   4515   & 0.87   & 2441  &0.874 & 2456  & 1.04  &1109  &$3.37\times10^{-4}$ &     16830&       2.05\\
2000 &   33.186 &   0.1104  &   4694   & 0.80   & 2317  &0.810 & 2678  & 0.97  &1126  &0.00352 &     18490&       1.90\\
4000 &   27.071 &   0.08661 &  4870    & 0.75   & 2221  &0.755 & 2915  & 0.89  &1140  &0.00938 &     20290&       1.75\\
8000 &   22.276 &   0.06855 &  5045    & 0.70   & 2122  &0.709 & 3163  & 0.82  &1151  &0.0170  &     22190&       1.61\\
10000 &  20.959 &   0.06370 &  5102    & 0.68   & 2096  &0.695 & 3244  & 0.80  &1154  &0.0197  &     22820&       1.57\\
%\hline
\end{tabular}
\end{ruledtabular}
\end{center}
\caption{Wavelengths $\lambda$ in \AA{} and absolute value of the
 oscillator strength in atomic units for a few of the lowest linear polarized
 transitions $\mu^{2S+1}0^+\rightarrow\nu^{2S+1}0^-$. The transition
 $1^10^+\rightarrow1^10^-$ is an example for a linear polarized
 transition involving a tightly bound state. For the transition
 $2^30^+\rightarrow1^30^-$ the oscillator strength shows a field
 dependence deviating from the typical behavior.
 }\label{tab:numval_lin}
\end{table*}

\begin{figure}
  \includegraphics[scale=0.4,angle=-90]{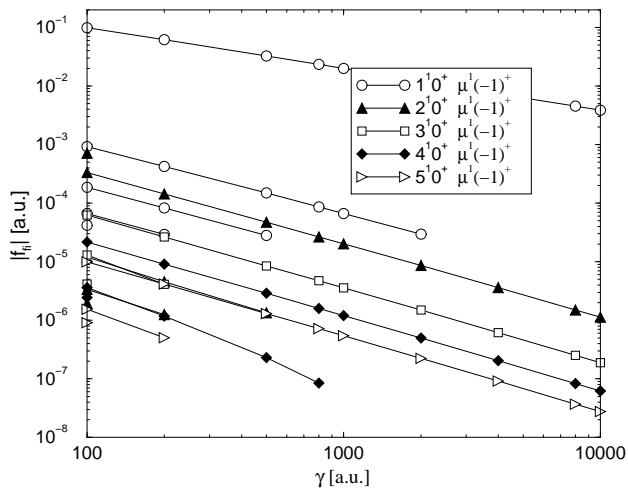}%
  \caption{\label{fig:osc_0+1+}The absolute value of the oscillator
    strength $|f_{fi}|$ of the circular polarized transitions 
    $\nu^10^+\rightarrow\mu^1(-1)^+$ as a function of the 
    field strength $\gamma$. On the left hand side from top to bottom the
    following transitions are shown $(\mu,\nu)=$ $(1,1)$,$(1,2)$,
    $(2,2)$, $(2,1)$, $(1,3)$, $(1,4)$, $(3,1)$, $(1,5)$, $(4,1)$, $(3,2)$,
    $(2,3)$, $(5,1)$, $(3,4)$, $(4,2)$, $(2,4)$, $(4,3)$, $(2,5)$,
    $(5,2)$, $(5,3)$.} 
\end{figure}

We present in  Fig.~\ref{fig:osc_0+1+} the oscillator strengths as a
function of the field strength for the circular polarized 
transitions of the form $\nu^10^+\rightarrow\mu^1(-1)^+$. 
 The typical power law dependence of the oscillator strengths is
 observed, as described in Sect.\ref{sec:electroMagTrans}.
 Furthermore for several 
transitions  we obtain $f_{fi}(\gamma)\approx C 
\gamma^{-\lambda}$ with a similar exponent $\lambda$,
i.e.
 parallel curves  on a double logarithmic scale. On the
other hand the reader observes that the number of transitions
decreases with increasing field strength,
being a consequence of the finite nuclear mass effects.  Transition wavelengths and oscillator strengths for the
transitions  $1^10^+\rightarrow1^1(-1)^+$,
$1^10^+\rightarrow2^1(-1)^+$, $1^30^+\rightarrow1^3(-1)^+$, and
$2^30^+\rightarrow1^3(-1)^+$ and oscillator strengths for the
transition $1^10^+\rightarrow1^1(-1)^+$ with finite mass effects excluded
are presented in table \ref{tab:numval_circ}.

\begin{table*}[htbp]
\begin{center}
\begin{ruledtabular}
\begin{tabular}{cddddcdcdc}
%\hline
 &  \multicolumn{3}{c}{$1^10^+\rightarrow1^1(-1)^+$}&\multicolumn{2}{c}{$1^10^+\rightarrow2^1(-1)^+$}&\multicolumn{2}{c}{$1^30^+\rightarrow1^3(-1)^+$}&\multicolumn{2}{c}{$2^30^+\rightarrow1^3(-1)^+$}\\
\cline{2-4}\cline{5-6}\cline{7-8}\cline{9-10}
\multicolumn{1}{c}{$\gamma$}     &
 \multicolumn{1}{c}{$\lambda(M_0,\gamma)$} &
 \multicolumn{1}{c}{$|f_{fi}(M_0,\gamma)|$}  &
 \multicolumn{1}{c}{$|f_{fi}(\infty,\gamma)|$}&
 \multicolumn{1}{c}{$\lambda(M_0,\gamma)$} &
 \multicolumn{1}{c}{$|f_{fi}(M_0,\gamma)|$} &
 \multicolumn{1}{c}{$\lambda(M_0,\gamma)$} &
 \multicolumn{1}{c}{$|f_{fi}(M_0,\gamma)|$} &
 \multicolumn{1}{c}{$\lambda(M_0,\gamma)$} &
 \multicolumn{1}{c}{$|f_{fi}(M_0,\gamma)|$}  \\
\hline
100 &    164.71  & 0.099  &     0.099   &       85.989&    $9.23\times10^{-4}$&  140.28 &  $4.04\times10^{-4}$          &       132.52 &$7.17\times10^{-5}$\\
200 &    135.56  & 0.062  &     0.061   &       67.898 &   $4.25\times10^{-4}$&  107.84 & $1.82\times10^{-4}$         &       102.93 &$3.16\times10^{-5}$\\
500 &    105.45  & 0.033  &     0.032   &       50.134 &   $1.489\times10^{-4}$&  77.711 &   $6.3\times10^{-5}$  &       74.960 &$1.08\times10^{-5}$\\
800 &    92.93   & 0.023  &     0.023   &       43.083 &    $8.6\times10^{-5}$&  66.278 &  $3.72\times10^{-5}$ &       64.204 &$6.2\times10^{-6}$ \\
1000 &   87.56   & 0.020  &    0.019    &       40.128 &   $6.66\times10^{-5}$&  61.590 & $2.9\times10^{-5}$   &       59.771 &$4.8\times10^{-6}$ \\
2000 &   72.77   & 0.012  &    0.012    &       32.285 &    $2.97\times10^{-5}$ &  49.521 &  $ 1.2\times10^{-5}$  &       48.290 &$2.1\times10^{-6}$\\
4000 &   60.25   & 0.0074 &    0.0068&&   &  40.494 &  $5.7\times10^{-6}$  &       39.636 & $9.2\times10^{-7}$\\
8000 &   49.31   & 0.0045  &   0.0040&&   &  33.843 &   $2.5\times10^{-6}$  &       33.222 & $4.0\times10^{-7}$\\
10000 &  46.02   & 0.0038  &   0.0033&&   &  32.137 &  $1.9\times10^{-6}$  &       31.571 &$3.0\times10^{-7}$ \\
%\hline
\end{tabular}
\end{ruledtabular}
\end{center}
\caption{Wavelengths $\lambda$ in \AA{} and absolute values of the
 oscillator strength in atomic units for the circular polarized
 transitons $1^10^+\rightarrow1^1(-1)^+$,
 $1^10^+\rightarrow2^1(-1)^+$, $1^30^+\rightarrow1^3(-1)^+$, and
 $2^30^+\rightarrow1^3(-1)^+$. Furthermore the table includes fixed
 nucleus results for the 
 oscillator strengths of the transition $1^10^+\rightarrow1^1(-1)^+$.}\label{tab:numval_circ}
\end{table*}

\begin{figure}
  \includegraphics[scale=0.4,angle=-90]{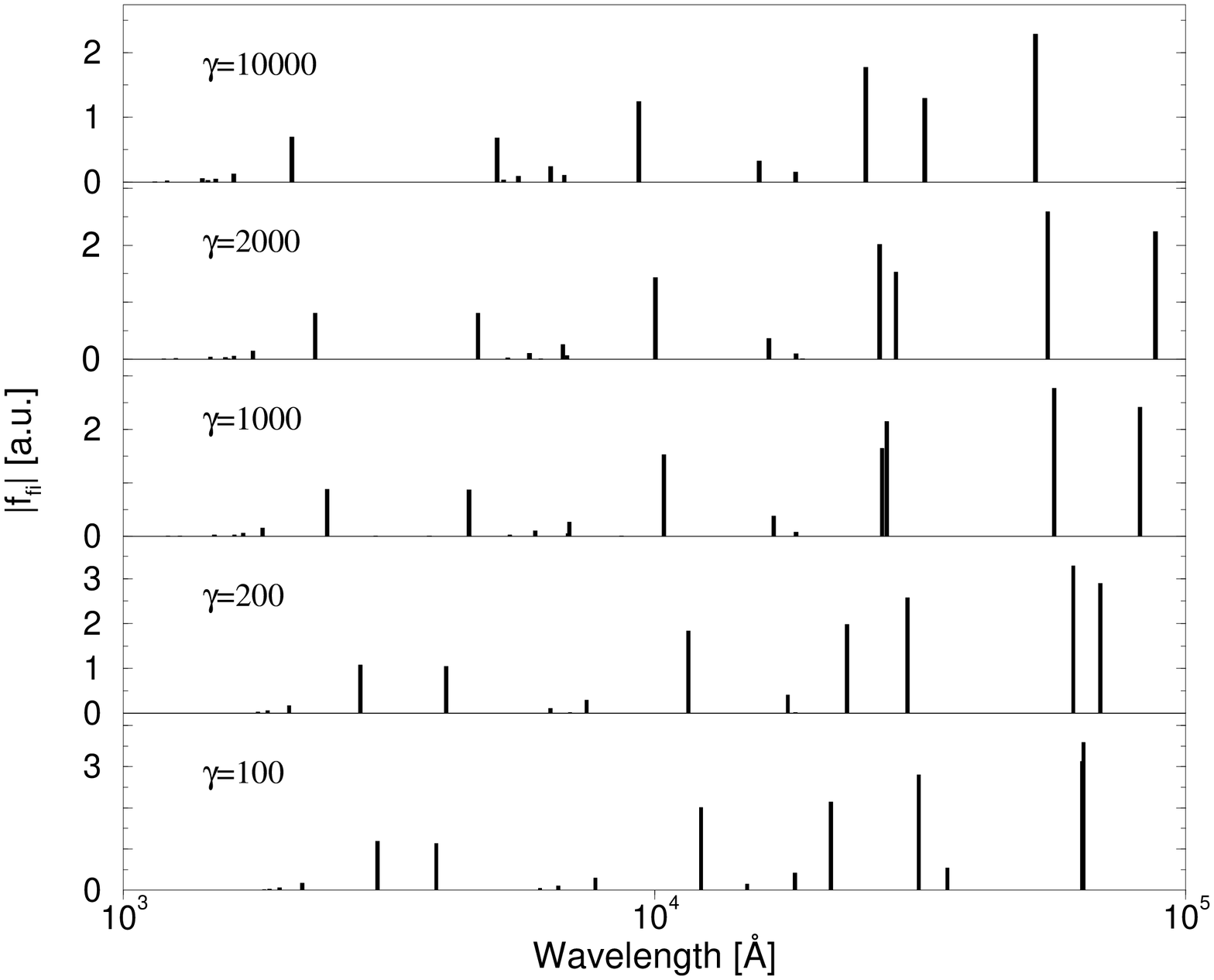}%
  \caption{\label{fig:osc_0+s}The oscillator strengths $|f_{fi}|$ of the
    linear and circular polarized transitions emanating 
    from the singlet states with zero magnetic quantum number and
    positive $z$~parity, i.e. $n^10^+,n=1,\ldots,5$ with their
    wavelength given in \AA{} for
    $\gamma=100,200,1000,2000,10000$ a.u. from bottom to top.}
\end{figure}

\begin{figure}
  \includegraphics[scale=0.4,angle=-90]{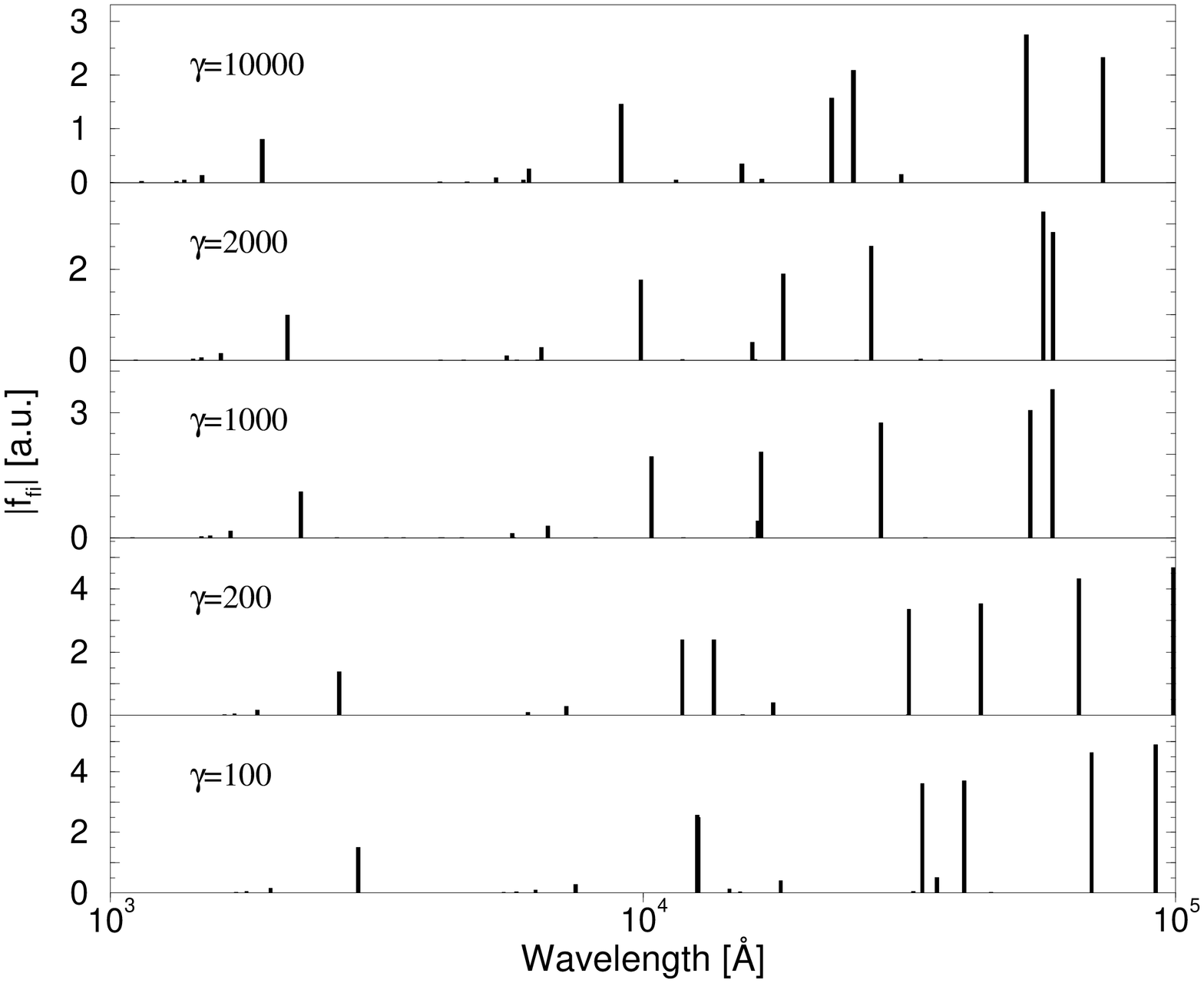}%
  \caption{\label{fig:osc_0+t}The oscillator strength $|f_{fi}|$ of the
    linear and circular polarized transitions emanating 
    from the triplet states with zero magnetic quantum number and
    positive $z$~parity, i.e. $n^30^+,n=1,\ldots,5$ with their
    wavelength given in \AA{} for
    $\gamma=100,200,1000,2000,10000$ a.u. from bottom to top.}
\end{figure}

\begin{figure}
  \includegraphics[scale=0.4,angle=-90]{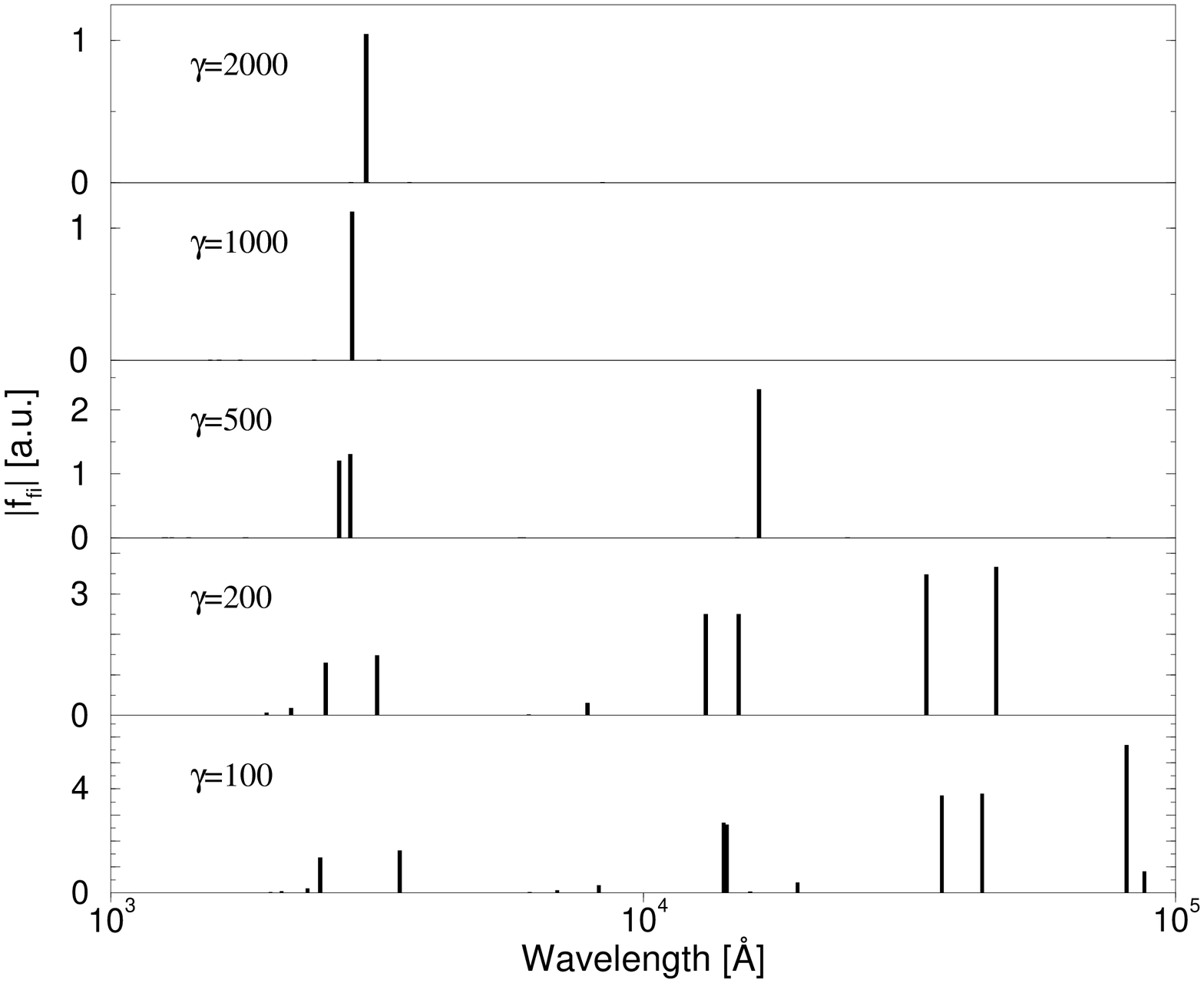}%
  \caption{\label{fig:osc_1-s}Same as in Fig.~\ref{fig:osc_0+s} but
    transitions emanating 
    from the singlet states with $n^1(-1)^-,n=1,\ldots,5$  for
    $\gamma=100,200,500,1000,2000$ a.u. from bottom to top.}
\end{figure}

\begin{figure}
  \includegraphics[scale=0.4,angle=-90]{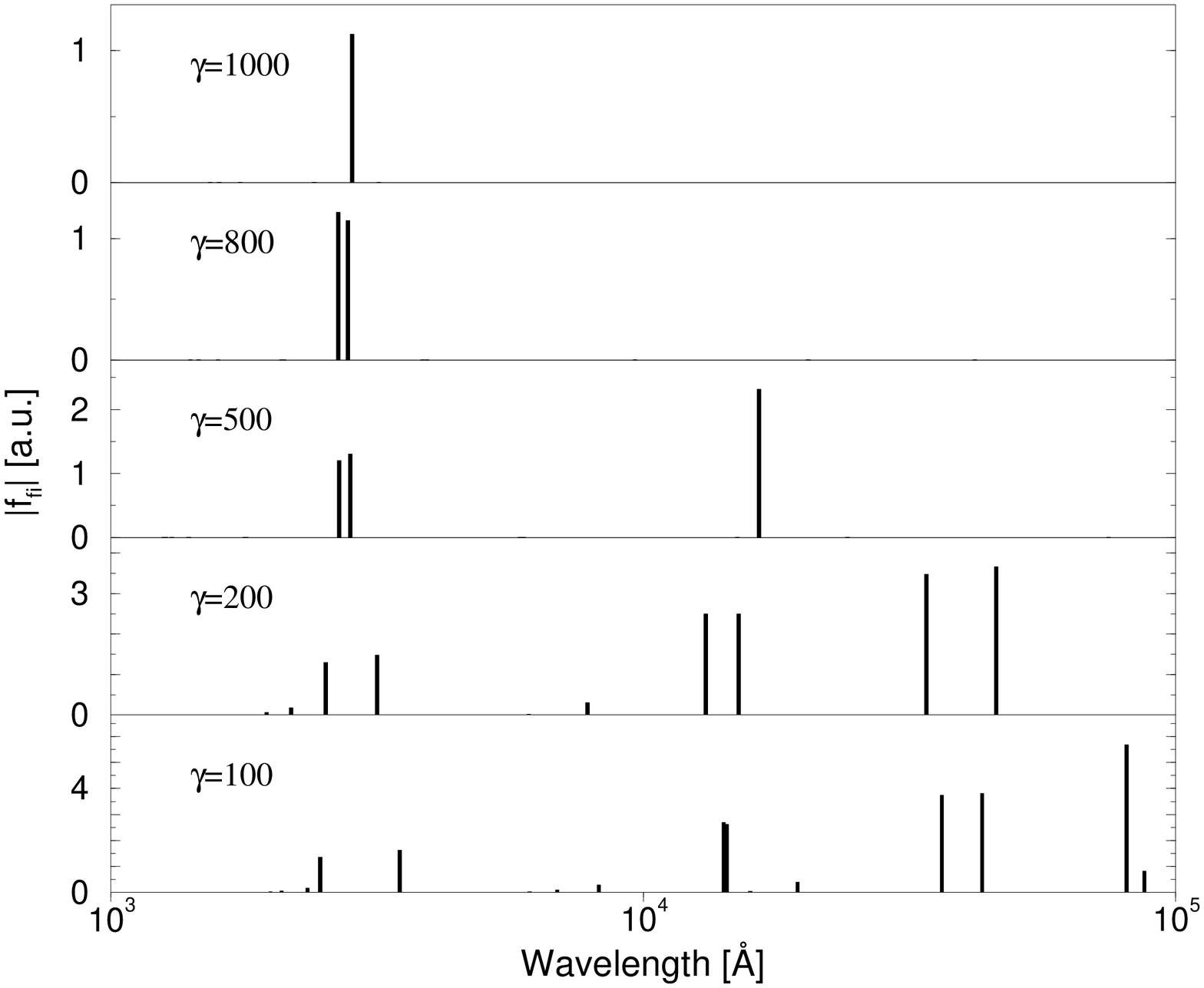}%
  \caption{\label{fig:osc_1-t} Same as in Fig.~\ref{fig:osc_1-s} but
    transitions emanating from the triplet states with magnetic
    quantum number $-1$ and negative $z$ parity for
    $\gamma=100,200,500,1000,2000$ a.u. from bottom to top.}
\end{figure}

\begin{figure}
  \includegraphics[scale=0.4,angle=-90]{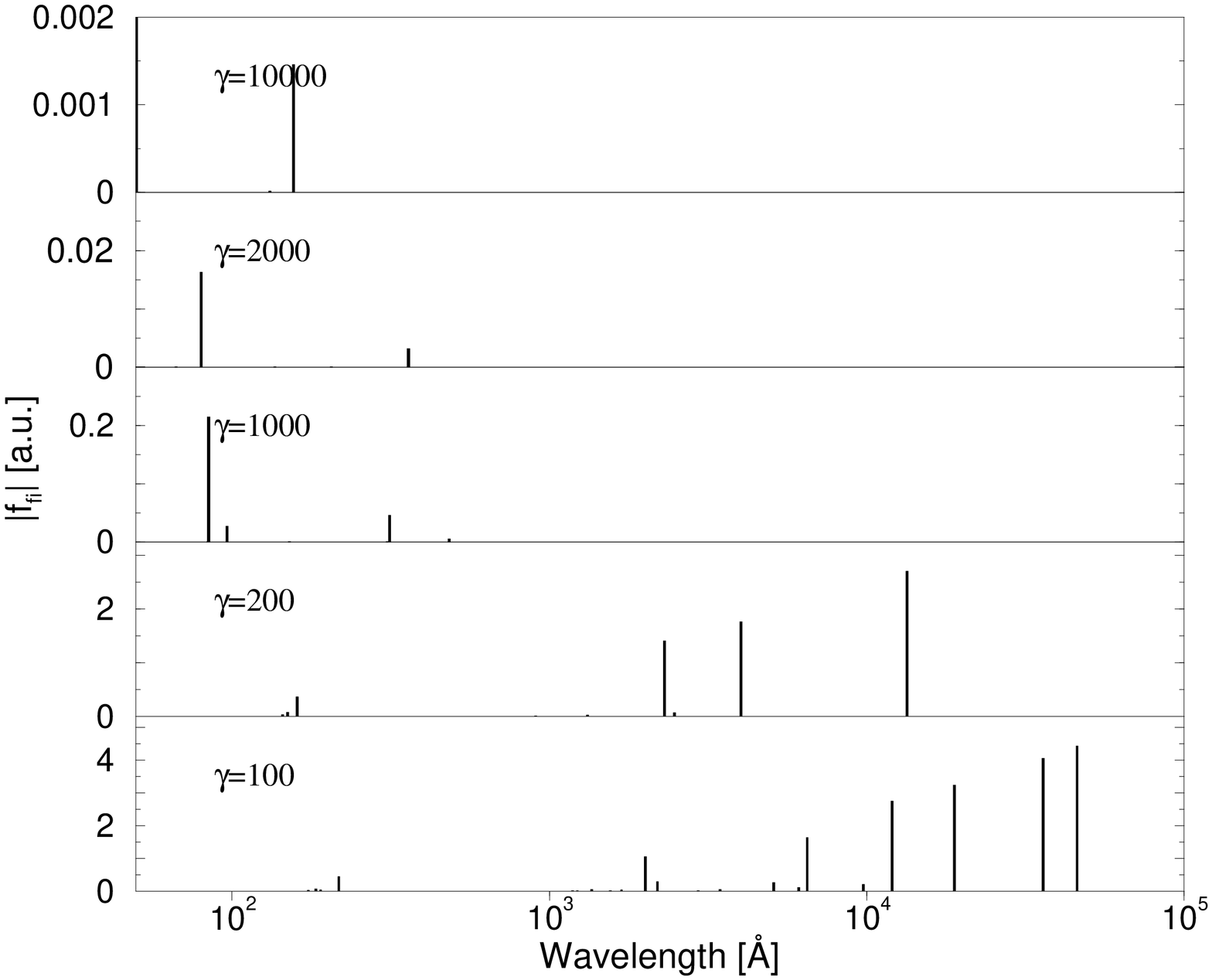}%
  \caption{\label{fig:osc_2+s}Same as in Fig.~\ref{fig:osc_0+s} but
    transitions emanating 
    from the singlet states with  magnetic quantum number $-2$ and
    positive $z$~parity, i.e. $n^1(-2)^+,n=1,\ldots,5$ with their
    wavelength given in \AA{} for
    $\gamma=100,200,1000,2000,10000$ a.u. from bottom to top. Note
    the different length scales on the oscillator strength axis, for
    different field strengths.} 
\end{figure}

\begin{figure}
  \includegraphics[scale=0.4,angle=-90]{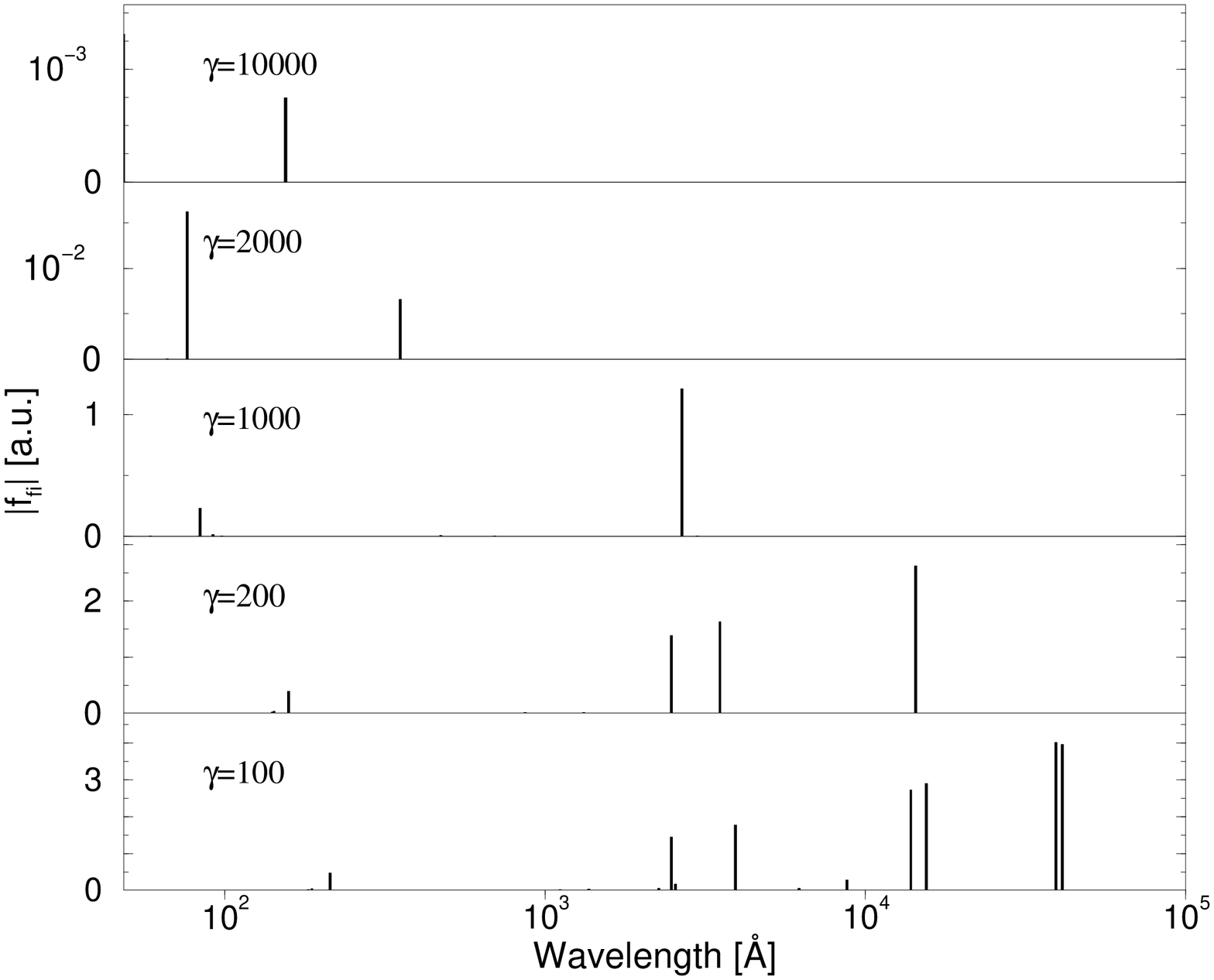}%
  \caption{\label{fig:osc_2+t}Same as in Fig.~\ref{fig:osc_2+s}, but
    for transitions of corresponding triplet states,
    i.e. $n^3(-2)^+,n=1,\ldots,5$ for 
    $\gamma=100,200,1000,2000,10000$ atomic units from bottom to top. Note
    the different length scales on the oscillator strength axis, for different field strengths.}
\end{figure}

In Figs.~\ref{fig:osc_0+s}~--~\ref{fig:osc_2+t} the oscillator strengths
of linear and circular polarized transitions are shown as a function of their
wavelengths for  different field strengths addressing the symmetry
subspaces $\nu^10^+$, $\nu^30^+$,
$\nu^1(-1)^-$, $\nu^3(-1)^-$, $\nu^1(-2)^+$, and $\nu^3(-2)^+$ for
$\nu=1,\ldots,5$. With the exception of Figs~\ref{fig:osc_2+s} and
\ref{fig:osc_2+t} the range of wavelength shown is $10^3$~\AA{}~--~$10^5$~\AA{}. 

Let us first discuss the oscillator strengths emanating from the
singlet states with zero magnetic quantum number and positive
$z$~parity (Fig.~\ref{fig:osc_0+s}). With increasing field strength
the transition wavelengths of some transitions decrease, whereas it
increases for others. E.g. at
$\approx3000$~\AA{} a gap between two groups of oscillator strengths
emerges and widens with increasing field strength. The reader should
note, that the values of the oscillator 
strengths correspondingly decrease.
Similar statements hold also for the spectrum of the triplet
transitions shown in 
Fig.~\ref{fig:osc_0+t}.

The transition spectrum  emanating from the states with magnetic
quantum number $-1$ and negative $z$~parity shows a completely
different pattern (see Figs.~\ref{fig:osc_1-s} and \ref{fig:osc_1-t}). The spectra are only
reported up to $\gamma=2000$, since above this field strength there
are no transitions between bound states including states of the $^{2S+1}(-1)^-$ symmetry.
It can be observed that at $\gamma=100$ there are several very
dominant transitions between $10^3$~\AA{} and $10^5$~\AA{} (up to 6 atomic
units). The largest of these disappear with increasing field strength
and at $\gamma=2000$ for $^1(-1)^-$, and 
$\gamma=1000$ for $^3(-1)^-$ respectively, only one
transitions with an oscillator strengths of the order of one remains.

In Figs.~\ref{fig:osc_2+s} and \ref{fig:osc_2+t}, the spectra for
transitions emanating from the $^{2S+1}(-2)^+$ symmetry subspaces, are
presented. For $\gamma=100,200$~a.u. the oscillator strength increase
(with a few exceptions) monotonically as a function of the wavelength. For
wavelengths of the order of $10^3$~\AA{}, we find only oscillator strengths 
much smaller than 1, whereas for wavelengths in the interval
$10^4-10^5$~\AA{} the corresponding quantities are in the range $3-5$.  At
$\gamma=10000$, only one 
transition of the order of $0.002$~atomic units for a wavelength of
approximately $10^2$~\AA{} remains.

\section{Brief conclusions}
\label{sec:concl}
We have applied a full configuration interaction method
to the helium atom in the superstrong   field regime between
$100$--$10\ 000$ atomic units. The effects of the finite nuclear mass
have been taken into  account. In this work we have presented results
on the oscillator strengths between bound states. The operators,
describing the dominating electric dipole transitions in the magnetic
field in first order
perturbation theory have been derived from first principles.
It has been shown how the spectrum changes for different symmetries
with increasing field strength. Finite nuclear mass effects decrease
the number of bound state transitions in the superstrong field regime,
since many states enter the continuum beyond a certain critical
field strength. 
The influence of the finite nuclear mass on the oscillator strength
has been analyzed. 

For linear polarized transitions that do not involve tightly bound
states the corresponding oscillator strengths are approximately
field-independent and, compared to the results for infinite nuclear
mass scaled by a factor involving the reduced mass. For linear
polarized transitions involving tightly bound states the oscillator
strengths obey a power law decay $f_{fi}(\gamma)\approx C
\gamma^{-\lambda}$. A similar statement holds for the circular
polarized transitions.  Particular linear and circular polarized
transitions do not belong to these two cases: they show a different
 strongly nonlinear  dependence on the field strength.

Our results could be of relevance to the interpretation of
 spectra of neutron stars. For the future an
investigation of the continuum would be very promising, particular
 since  the discrete spectrum becomes very
sparse in a superstrong field. The inclusion of motional electric
 fields into our study would also be very desirable. The latter
 requires however a major theoretical effort.

\section{Acknowledgments}
\label{sec:acknol}
The Deutsche Forschungsgemeinschaft is gratefully acknowledged for
financial support. Discussions with S. Jordan, D. Wickramasinghe and
J. Liebert are gratefully acknowledged.

%\bibliography{Astrophysik_Literatur,H_Literatur,He_Literatur_finite,He_Literatur_inifinte,Wasserstoff,Sonstige,Hminus}

\end{document}